\documentclass[a4paper,10pt]{article}
\usepackage{graphicx}

\begin{document}

\setlength{\oddsidemargin} {1cm}
\setlength{\textwidth}{18cm}
\setlength{\textheight}{23cm}

\title{Study of chaotic motion in fluid mechanics with the scale relativity methods}

\author{{Marie-No\"elle C\'el\'erier} \\
{\small Laboratoire Univers et TH\'eories (LUTH), Observatoire de Paris, CNRS \& Universit\'e Paris Diderot} \\
{\small 5 place Jules Janssen, 92190 Meudon, France} \\
{\small e-mail: marie-noelle.celerier@obspm.fr}}

\maketitle

\begin{abstract}

Chaotic motion in time of a number of macroscopic systems has been analyzed, in the framework of scale relativity, as motion in a fractal space with topological dimension 3 and geodesics with fractal dimension 2. The motion equation is then Schr\"odinger-like and its interpretation in fluid mechanics gives the well-known Euler and Navier-Stokes equations. We generalize here this formalism to the study of a system exhibiting a chaotic behavior both in space and time. We are thus lead to consider macroscopic fluid properties as issuing from the geodesic features of a fractal `space-time' with topological dimension 4 and geodesics with fractal dimension 2. This allows us to obtain both a motion equation for the fluid velocity field, which exhibits then three components while only one is necessary for the description of an ordinary fluid, and a relation between their three curls. The physical properties of this solution suggest it could represent some three-dimensional chaotic behavior for a classical fluid, tentatively turbulent if particular conditions are fulfilled. Different ways of testing experimentally these assumptions are proposed.

\end{abstract}

\section{Introduction}

The first transformations of the quantum mechanics equations into fluid mechanics-type equations lead to unsuccessful interpretation attempts. The decomposition of the Schr\"odinger equation in its real and imaginary parts actually allows one to obtain Euler and continuity-like equations, but the appearance in the Euler equation of a ``quantum potential'' with an amplitude proportional to $\hbar^2$ and the {\em axiomatic and a posteriori} interpretation of the phase gradient as a velocity did not allow the physicists of the time to propose an acceptable interpretation of what appeared as a mathematical exercise without any physical application \cite{EM27,DB52,TT52}.

The same method applied to the Klein-Gordon equation gave results still more unexplainable in the framework of the physics known in those days \cite{TT52}. These transformations were thus relegated to the cupboard of the false good ideas for half a century.

A significant improvement was made however with the appearance, in the framework of scale relativity, of a Schr\"odinger-like equation generalized to the macroscopic domain \cite{LN93}. Here the quantum mechanical coefficient $\hbar/2m$ is identified with a macroscopic constant, ${\cal D}$, characteristic of the system under study and which appears in the expression for the quantum potential, itself become macroscopic. In this interpretation, this quantum potential stems from the fractal geometry of space(-time) and the velocity field of the physical fluid is identified to that of the geodesic fluid. Note, however, ${\cal D}$ does not characterize space-time directly but its geodesics. It is therefore possible to have only one space-time including different geodesic bundles whose various properties correspond to different physical systems \cite{LN09a}.

Various particular cases of transformations of the Schr\"odinger equation, some direct, some inverse, into fluid mechanics equations have been studied in a scale relativity framework \cite{LN09a}:

\begin{itemize}

\item Integration et combination of the Euler and continuity equations with a quantum potential allowing one to obtain a linear (without pressure) or a non-linear (with pressure) Schr\"odinger equation for an irrotational fluid, and inverse transformations.

\item Fluid motion with vorticity, hence possible turbulence, corresponding to a Schr\"odinger equation in which vorticity plays a role similar to that of an external magnetic field.

\item Motion equation for a charged fluid submitted to an electromagnetic field and a quantum potential equivalent to a Ginzburg-Landau-type equation for superconductivity. Such a fluid could thus exhibit some properties of a quantum fluid.

\item The same method has been applied in a formal way to the Navier-Stockes equation. Here the viscosity term is given a role similar to that of the ${\cal D}$ parameter \cite{LN09b}.

\end{itemize}

An equivalent transformation has been applied to the free Klein-Gordon equation. It gives an Euler-type equation for a relativistic fluid and a conservation law for a relativistic ``current'' \cite{LN05}. The main steps of this derivation are recalled in the next Section.

Applying this technique to a free quaternionic Klein-Gordon equation allows one to go further on and to characterize chaotic motion, and possibly some kind of turbulent behavior, for a macroscopic fluid. We display in this article, the main steps leading to these results. The interested readers can find the details of the reasonings and calculations in \cite{MNC09}.

\section{Relativistic Euler equation}

Following the M\"adelung transformation, the technique consists in separating the real and pure imaginary parts in the motion equation which, in the simplest case (spin-less particle), is a standard Klein-Gordon equation, i.e. written with complex numbers.

The application of the usual method used by M\"adelung \cite{EM27} and followers leads to obtain, from the real part, a Hamilton-Jacobi-type equation which must by differentiated to yield the Euler equation and, from the pure imaginary part, directly the continuity equation.

With the scale relativity methods, the calculations on the real part leading to the Euler equation can be simplified. Just remember that, in this framework, the Klein-Gordon equation is obtained by integrating a geodesic equation. It is therefore simpler to avoid integrating then differentiating again the real part and to only extract directly this real part from the geodesic equation before integration. This yields an Euler-type equation with a macroscopic quantum potential, both relativistic.

It is easy to check that, at the non-relativistic limit, the usual Euler equation with quantum potential is recovered.

However, to obtain the continuity equation, it is mandatory to go through the integration of the geodesic equation into a Klein-Gordon equation. One then extracts its pure imaginary part. At the non-relativistic limit, the continuity equation reads
\begin{equation}
\frac{\partial P}{\partial t} + \nabla(PV) = 0.
\end{equation}
The squared modulus of the wave function, $P$, can therefore be interpreted as a density, $\rho = mP$, for the fluid of the fractal space-time geodesics. Within this interpretation, the geodesic fluid is more concentrated in some places than in others. It completely fills some space-time regions and others are nearly empty, as with a fluid \cite{NC07}.

\section{Quaternionic Klein-Gordon equation}

The system we are going to study now is a fluid which is chaotic both in space and time. Our hope is that this fluid might be able to represent some type of turbulence which double chaotic nature has been experimentally verified a number of times.

The use of a macroscopic quantum-like equation is justified here by this chaotic nature of the fluid motion. It is well-known that in non-(motion)relativistic scale relativity the equation describing a macroscopic system with chaotic time evolution is Schr\"odinger-like. For chaotic motion both in space and time, a generalization to motion described in a four-dimensional space-time, scale-dependent both in space and time, should therefore constitute an a priori valid approach.

First, recall that, in quantum mechanics, the free Dirac equation is the square root of the free Klein-Gordon equation written with bi-quaternions (complex quaternions). The use of bi-quaternions is imposed by the breakings of the symmetries ${\rm d}s \leftrightarrow - \, {\rm d}s$, ${\rm d}x^{\mu} \leftrightarrow -\, {\rm d}x^{\mu}$ and $x^{\mu} \leftrightarrow -\, x^{\mu}$ \cite{CN04}. Now, in macroscopic mechanics, the breaking of the discrete symmetry $x^{\mu} \leftrightarrow -\, x^{\mu}$, parity and time reversal, is not a property of the physical laws. We therefore need only use in this case a merely quaternionic geodesic equation \footnote{Remember quaternions are generalized complex numbers with four components, one real and three imaginaries, whose product possesses the particular property to be non-commutative: $A \times B \neq B \times A$.}.

It should not be justified to claim the macroscopic fluid which will emerge from this study will exhibit every quantum-type property, nor will it be relativistic. This is forbidden by the non-physical nature of the quaternionic Klein-Gordon equation unable to yield a Dirac equation for bi-spinors which should be the only acceptable one. We must consider the quaternionic construction presented here as a representation of some kind of motion chaotic both in space and time, only valid at the non-relativistic limit and in the macroscopic domain.

The mathematical construction is analogous to that developed for quantum mechanics,
save that it does not depend on $\hbar$, but on the macroscopic parameter ${\cal D}$ which characterizes the transition scale from scale dependence to scale independence. It follows the different steps displayed hereafter.

\begin{itemize}

\item We introduce a ``quaternionic wave function'' which is a mere redefinition of a quaternionic action.

\item It allows us to define a quaternionic velocity field whose four components (the real one and the three pure imaginary parts) are written parametrically.   

\item We find that three fields (the real one and two among the pure imaginary components) have non-vanishing curls which are linked together by a set of three differential equations.

\item The covariant derivative applied to the quaternionic velocity gives a geodesic equation whose real part is an Euler-type equation with a relativistic quantum potential.

\item We integrate the geodesic equation previously found.

\item We extract the three pure imaginary parts of the resulting equation which give three continuity equations, one for each vectorial component of the velocity field.

\item We take the non-relativistic limit of each of the seven equations thus found (Euler, continuity and curls).

\end{itemize}

\section{Results}

\subsection{General results}

An examination of the seven equations eventually obtained allows us to conclude that the properties which apply to the most general fluids able to be described with this formalism are as follows:

\begin{itemize}

\item an Euler-type motion equation for a ``3-fluid'' with vorticity, corresponding to a fluid with three velocity-type components, each being a 3-vector with spatial components.

\item the same quantum potential as the one obtained with the transformation of the Schr\"odinger equation, i.e. proportional to ${\cal D}^2$.

\item the curls of the three velocity fields are non-vanishing and perpendicular ones with the others. Moreover, their amplitudes are inversely proportional to ${\cal D}$ and hence to the strength of the potential.

\item ${\bf {\cal D}}$ has the dimension of a diffusion coefficient.

\end{itemize}

\subsection{Particular case: Beltrami flow}

When the three velocity-components of the fluid verify
\begin{equation}
v_i = v_j \times v_k,
\end{equation}
the solutions are Beltrami flow-type. This sort of flow is actually characterized by a property of its velocity curl which happens to be parallel to this velocity.

Given that the Beltrami flows are chaotic, we deduce that the general solutions described above are at the very least chaotic, since they include Beltrami as particular cases.

\subsection{Turbulence}

It has been experimentally established that turbulence appears in a fluid with kinematic viscosity coefficient $\nu$, when its Reynolds number, $Re \equiv V \, L/ \nu$ (where $V$ et $L$ are velocities and lengths characteristic of the flow) becomes larger than some critical value:
\begin{equation}
Re_c \equiv V_t \, L_t/ \nu.
\end{equation}

Now, in our model, the transition from chaotic to non-chaotic flow is characterized by the ${\cal D}$ parameter which is a relative quantity, specific to every system, and of which an order of magnitude can be given by a generalized Heisenberg relation: $\delta x \, \delta v \simeq 2 {\cal D}$, where $\delta v$ is a velocity dispersion and, $\delta x$, a position dispersion \cite{DD04}. A critical Reynolds number might thus be associated to $2 {\cal D} = V_t L_t$, where $V_t L_t$ $= Re_c \nu = 2 {\cal D}$. If such were the case, it could be an indication that some turbulent-type flow might be represented by a velocity field with three components analogous to that previously described.

\subsection{Spin and vorticity}

The spontaneous appearance of vorticity at a macroscopic level for a chaotic behavior in space and time can be compared to that of spin in quantum mechanics. Both derive from the fractal nature of space-time.

It is worth stressing however both do not emerge at the same construction level of the theory. The spinor corresponds to the wave function, while the chaotic vorticity corresponds to its derivative.

\section{Experimental test proposals}

The generalized chaotic behavior obtained here can be considered as the fondamental state emerging from the total giving up of the differentiability assumption which leads to the breakings of the symmetries ${\rm d}s \leftrightarrow - \, {\rm d}s$ {\it and} ${\rm d}x^{\mu} \leftrightarrow - \, {\rm d}x^{\mu}$. Chaotic motion in time only is the degenerate state where only the symmetry ${\rm d}t \leftrightarrow - \, {\rm d}t$ is broken. It gives, in the inertial case, which is the simpler one and to which we have limited our study here, a macroscopic Schr\"odinger-like equation from which one can derive an Euler equation for an irrotational hence non chaotic fluid.

However, there is no a priori reason for the parameter ${\cal D}_l$ yielding the Schr\"odinger equation from which a laminar behavior occurs to have the same value as the parameter ${\cal D}_t$ giving a Klein-Gordon equation from which a chaotic behavior can emerge, even for the same fluid. We might suspect that the length scale related to the former is of the order of the mean free path of the fluid molecules, while we know from experiment that the transition to chaos occurs at much larger scales. Therefore ${\cal D}_t$ could be considered as defining the scales characterizing the transition from non-chaotic to chaotic behavior, and possibly, the critical Reynolds number of the fluid.

We have also shown that ${\cal D}_t$ appears both in the ``quantum'' potential expression and in the curls of the velocity field components.

All this allows us to propose three means of testing our results.

A first test should be first to measure or fix ${\cal D}$ in the chaotic phase of a given fluid subjected to a quantum potential proportional to ${\cal D}^2$, then to verify if the transition from a chaotic to a laminar behavior occurs around some critical Reynolds number $Re_c$ associated to this value of ${\cal D}$.

A second test proposal is linked to our result that the amplitude of the velocity component curls are inversely proportional to ${\cal D}$: the greater the ``quantum'' or dissipative potential applied to the fluid, the less whirling and therefore the less chaotic it might be. This seems consistent with the observation that spontaneous turbulence disappears when viscous dissipation erases it. However, it would deserve a more exact and quantitative experimental confirmation.

The last test would be to examine if the velocity component curls are indeed inversely proportional to some ``critical Reynolds number'' multiplied by the fluid kinematic coefficient.

\section{Conclusion}

The scale relativity methods have been used to obtain the representation of a chaotic fluid under the form of a rotational 3-fluid.

Its motion equations include a ``quantum-type'' macroscopic potential  whose magnitude is determined by the chaotic/non-chaotic transition scale.

The curls of the three velocity fields of the 3-fluid are mutually perpendicular and inversely proportional to this scale.

We have proposed three different experiments able to test if this model could represent some kind of turbulence. Given the difficulties encountered in physics to model turbulent behaviors in fluids, it should be of importance that such experiences might be realized in a near future.

Now, we want to stress that the rather simple models for chaotic motion obtained here stem from a very preliminary approach. Richer properties are expected from more complex constructions which might be realized in future works.

\end{document}